**Piotr GAS**

AGH University of Science and Technology

# Temperature Distribution of Human Tissue in Interstitial Microwave Hyperthermia

***Abstract**. A model which is an example of local interstitial microwave hyperthermia is presented. A microwave coaxial-slot antenna placed in the liver tissue is the heat source. Due to the axial symmetry of the model, for simplification a two-dimensional case is considered. The presented issue is therefore a coupling of the electromagnetic field and the temperature field. Using the finite element method, the wave equation for TM wave case and the bioheat equation under steady-state condition have been solved. At the end the obtained simulation results are presented.*

***Streszczenie**. W pracy przedstawiono 2D model będący przykładem zastosowania miejscowej hipertermii śródmiąższowej. Źródłem ciepła jest współosiowa antena mikrofalowa z szczeliną powietrzną umieszczona w tkance wątroby. Przedstawiony problem stanowi sprzężenie pola elektromagnetycznego i pola temperatury. Posługując się MES rozwiązano równanie falowe dla przypadku fali TM, a następnie biologiczne równanie ciepła w przypadku stacjonarnym. Na końcu przedstawiono uzyskane wyniki symulacji. **(Rozkład temperatury tkanki ludzkiej w śródmiąższowej hipertermii mikrofalowej)***

## Introduction

Hyperthermia is a method of therapy in which the pathological tissues are exposed to high temperatures exceeding $40^{\circ}\mathrm{C}$. There are many techniques for hyperthermia treatment but interstitial hyperthermia seems to be the most effective one, because it delivers the heat directly at the site of the tumor and minimally affects surrounding the normal tissues. For interstitial hyperthermia high frequency needle electrodes, microwave antennas, ultrasound transducers, laser fibre optic conductors, or ferromagnetic rods, seeds or fluids are injected or implanted into the tumor [1, 2]. With these applicators high enough heat can be applied to induce thermonecrosis in tumors and cancerous tissues (which are located deep within the human body) at the distance of 1 to 2 cm around the heat source. This technique is suitable for tumors less than 5 cm in diameter [3]. Moreover, interstitial hyperthermia has reached positive clinical results in combination with radiotherapy and chemotherapy. This method is now gaining new fields of applications, for instance in the treatment of liver, breast, kidney, bone and lung tumors [4]. There are many studies on the treatment of cancer using hyperthermia which demonstrate that this aspect is still important and more research is needed in this matter [5, 6]. Scientists are still looking for new techniques that will make hyperthermia a simpler, safer, more effective and widely available method for patients. The use of nanotechnology in hyperthermia treatment e.g., magnetic fluid hyperthermia, which is currently under experimentation, seems particularly promising [7]. Historical background of the hyperthermia treatment can be found in [8].

In this article temperature distribution in the liver tissue, induced by the coaxial-slot antenna with defined the total input power, is calculated.

## Main equations and model geometry

Let us consider the coaxial – slot antenna as shown in Figures 1 – 2. Due to the axial symmetry the cylindrical coordinates $r$, $z$, $\phi$ are used. The antenna consists of an inner conductor, dielectric, outer conductor, and plastic catheter. The last one is a sleeve for the inner parts of the antenna. There are also two symmetric air slots in the external conductor. Each of them has the size $d$. Antenna dimensions are given in Table 1. The computational domain is in reality larger than indicated in Fig. 2. Because the model is axial-symmetric model, only a half of the geometry structure of the antenna and its surrounding human tissue are considered. The whole 3D model of the antenna immersed in the liver tissue will be created when the adopted 2D model rotates along the $z$ axis at $r = 0$.

Let us start with the Maxwell-Ampere's and Faraday's laws in the frequency domain:

$$\nabla \times \mathbf{H} = \mathbf{J} + \mathrm{j}\omega \mathbf{D} \tag{1}$$

$$\nabla \times \mathbf{E} = -\mathrm{j}\omega \mathbf{B} \tag{2}$$

where $\mathbf{E}$ and $\mathbf{H}$ are the electric and magnetic field strengths respectively, $\omega$ is the angular frequency of the electromagnetic field and $\mathbf{J}$ is the current density, which in conducting media is given by Ohm's law

$$\mathbf{J} = \sigma \mathbf{E} \tag{3}$$

where $\sigma$ is the electrical conductivity of the body. Moreover $\mathbf{D}$ and $\mathbf{B}$ are respectively the vectors of electric displacement density and magnetic induction given in the form of material dependences

$$\mathbf{D} = \varepsilon \mathbf{E} = \varepsilon_{\mathrm{r}} \varepsilon_0 \mathbf{E} \tag{4}$$

$$\mathbf{B} = \mu \mathbf{H} = \mu_{\mathrm{r}} \mu_0 \mathbf{H} \tag{5}$$

where $\varepsilon$ and $\mu$ are the permittivity and magnetic permeability of the medium respectively.

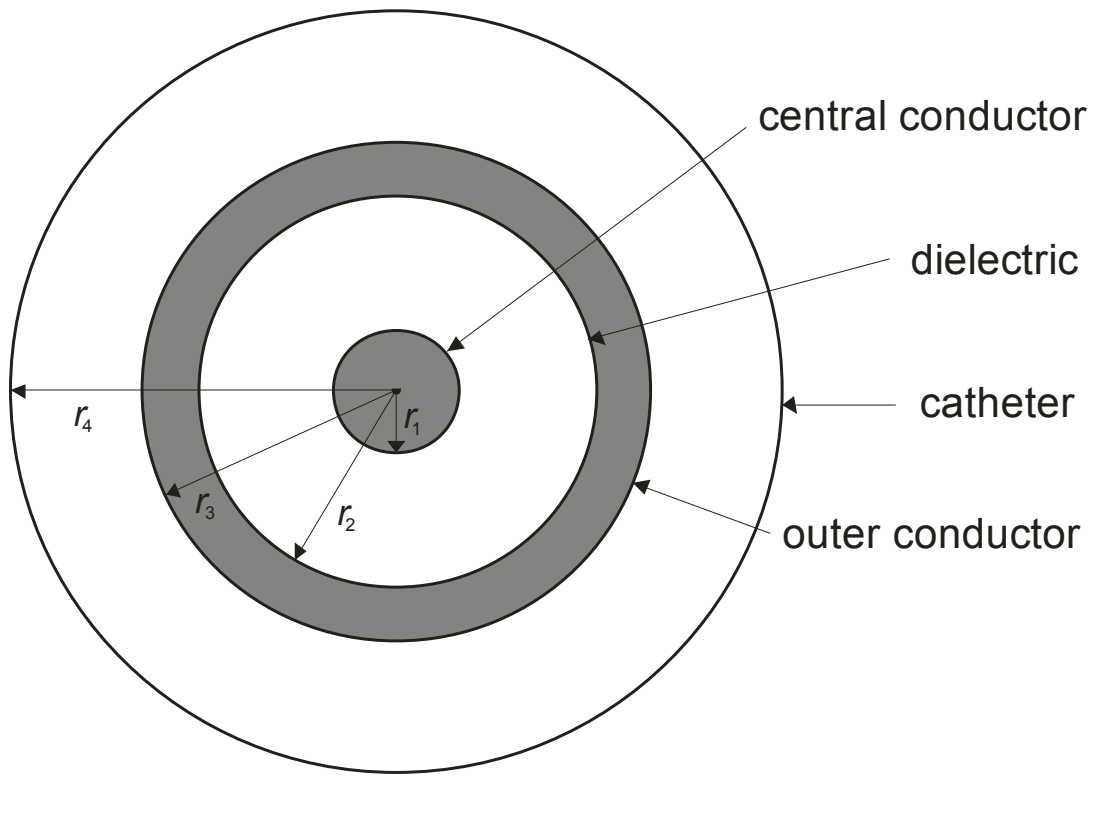

Fig.1. Cross section of the antenna with geometrical dimensions

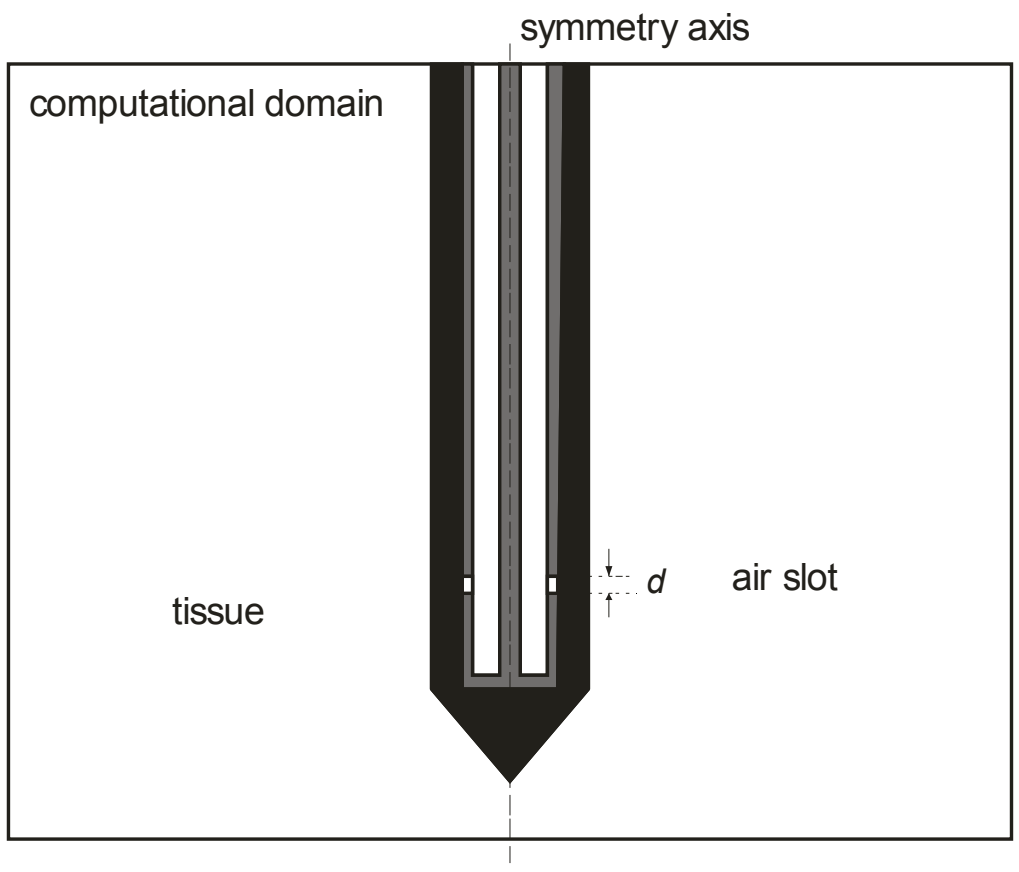

Fig.2. Schematic view of the part of the coaxial antenna with the air slots located in the human tissue

Table 1. Geometrical dimensions of the antenna in [mm]

| | |
|---|---|
| radius of the central conductor | $r_1$ = 0.135 |
| inner radius of the outer conductor | $r_2$ = 0.470 |
| outer radius of the outer conductor | $r_3$ = 0.595 |
| radius of the catheter | $r_4$ = 0.895 |
| size of the air slot | $d$ = 1 |

After taking into account the above dependences (3) to (5), the Maxwell's equations (1) and (2) take the following form

(6) $$\nabla\times\mathbf{H} = \mathrm{j}\omega\varepsilon_0\left(\varepsilon_r - \mathrm{j}\frac{\sigma}{\varepsilon_0\omega}\right)\mathbf{E}$$

(7) $$\nabla\times\mathbf{E} = -\mathrm{j}\omega\mu_\mathrm{r}\mu_0\mathbf{H}$$

where $\varepsilon_\mathrm{r}$ and $\mu_\mathrm{r}$ are the relative permittivity and relative permeability of the medium, $\varepsilon_0$ and $\mu_0$ are the permittivity and permeability of the vacuum, respectively. After applying the rotation operator to both sides of equation (6) and substituting equation (7) into (6) we can derive the following equation describing the field distribution:

(8) $$\nabla\times\left[\left(\varepsilon_\mathrm{r} - \mathrm{j}\frac{\sigma}{\omega\varepsilon_0}\right)^{-1}\nabla\times\mathbf{H}\right] - \mu_\mathrm{r}k_0^2\mathbf{H} = 0$$

where $k_0$ is the wave number in vacuum defined as

(9) $$k_0 = \omega\sqrt{\varepsilon_0\mu_0}$$

Since the vector of magnetic field strength (as well as $\mathbf{E}$) in the time domain is related with complex amplitude by

(10) $$\mathbf{H}(\mathbf{r},t) = \mathrm{Re}\left[\hat{\mathbf{H}}(\mathbf{r})\,\mathrm{e}^{\mathrm{j}\omega t}\right]$$

therefore equation (8) in the complex domain is given by

(11) $$\nabla\times\left[\hat{\varepsilon}_\mathrm{r}^{-1}\nabla\times\hat{\mathbf{H}}\right] - \mu_\mathrm{r}k_0^2\hat{\mathbf{H}} = 0$$

where $\hat{\varepsilon}_r$ is the complex relative permittivity defined by

(12) $$\hat{\varepsilon}_\mathrm{r}(\omega) = \varepsilon_\mathrm{r} - \mathrm{j}\frac{\sigma}{\omega\varepsilon_0}$$

Also the vector of electric field strength has a complex value which is determined from the relation

(13) $$\nabla\times\hat{\mathbf{E}} = -\mathrm{j}\omega\mu_\mathrm{r}\mu_0\hat{\mathbf{H}}$$

In the presented model transverse magnetic (TM) waves are used and there are no electromagnetic field variations in the azimuthal direction. A magnetic field $\mathbf{H}$ has only the $\phi$ -component and an electric field $\mathbf{E}$ propagates in $r$-$z$ plane. Therefore, it can be written as:

(14) $$\hat{\mathbf{H}}(\mathbf{r},t) = H_\phi(r,z)\mathbf{e}_\phi\,\mathrm{e}^{\mathrm{j}\omega t}$$

(15) $$\hat{\mathbf{E}}(\mathbf{r},t) = \left[E_r(r,z)\mathbf{e}_r + E_z(r,z)\mathbf{e}_z\right]\mathrm{e}^{\mathrm{j}\omega t}$$

In the axial-symmetric mode, the wave equation takes the form of scalar equation as follows

(16) $$\nabla\times\left[\left(\varepsilon_\mathrm{r} - \mathrm{j}\frac{\sigma}{\omega\varepsilon_0}\right)^{-1}\nabla\times H_\phi\right] - \mu_\mathrm{r}k_0^2H_\phi = 0$$

The presented problem models the metallic parts of the antenna using boundary conditions. For all metallic surfaces, the PEC (perfect electric conductor) boundary conditions are set as

(17) $$\mathbf{n}\times\mathbf{E} = 0$$

The external boundaries of the computational domain, which do not represent a physical boundary (except the boundary at the $z$ symmetry axis where $E_\phi(r)$ = 0) have the so-called matched boundary conditions. They make the boundary totally non-reflecting and assume the form

(18) $$\sqrt{\varepsilon - \mathrm{j}\frac{\sigma}{\omega}}\,\mathbf{n}\times\mathbf{E} - \sqrt{\mu}H_\phi = -2\sqrt{\mu}H_{\phi0}$$

where $\boldsymbol{H}_{\phi0}$ is an input field incident on the antenna given by

(19) $$H_{\phi0} = \frac{1}{Zr}\sqrt{\frac{Z P_\mathrm{in}}{\pi\ln(r_2/r_1)}}$$

In above equation $P_\mathrm{in}$ is the total input power in dielectric, while $r_1$ and $r_2$ are the dielectric's inner and outer radii, respectively. Moreover, Z signifies the wave impedance of the dielectric which is defined as

(20) $$Z = \sqrt{\frac{\mu_0}{\varepsilon_0\varepsilon_\mathrm{r}}}$$

The seed point is modelled using a port boundary condition with the power level set to $P_\mathrm{in}$ = 3W at the low-reflection external boundary of the coaxial dielectric cable.

The second basic equation used in the presented simulation is the so-called bioheat equation given by Pennes in the mid-twentieth century [9]. It describes the phenomenon of transport and heat transfer in biological tissues. In steady-state analysis it is expressed by

(21) $$\nabla(-k\nabla T) = \rho_\mathrm{b}C_\mathrm{b}\omega_\mathrm{b}(T_\mathrm{b} - T) + Q_\mathrm{ext} + Q_\mathrm{met}$$

where $T$ is body temperature [K], $k$ − tissue thermal conductivity [W/(m$^2$ K)], $T_\mathrm{b}$ − blood vessel temperature [K], $\rho_\mathrm{b}$ − blood density [kg/m$^3$], $\omega_\mathrm{b}$ − blood perfusion rate [1/s], $C_\mathrm{b}$ − blood specific heat [J/(kg K)]. The described model takes into account both the metabolic heat generation rate $Q_\mathrm{met}$ [W/m$^3$] as well as the external heat sources $Q_\mathrm{ext}$ [W/m$^3$], which is responsible for the changing of the temperature inside the exposed body according to the following equation

$$Q_{\text{ext}} = \frac{1}{2}\sigma\,\hat{\mathbf{E}}\cdot\hat{\mathbf{E}}^{*} = \frac{1}{2}\sigma\left|\hat{\mathbf{E}}\right|^{2} \tag{22}$$

The boundary condition explaining heat exchange on the surface of the human tissue uses insulation and is given as

$$\mathbf{n}\cdot\left(k\nabla T\right) = 0 \tag{23}$$

where $\mathbf{n}$ is the unit vector normal to the surface.

## Simulation results

In the analyzed model, the liver tissue and antenna are considered as homogeneous media with averaged material parameters. The antenna operates at the frequency $f = 2.45\ \text{GHz}$. Moreover, blood parameters are given in Tab. 2 and the metabolic heat generation rate of the tissue is set as $Q_{\text{met}} = 300$ [W/m$^3$]. Moreover, the tissue thermal conductivity is equal to $k = 0.56$ [W/(m$^2$ K)]. Other electrical parameters of the model are given in Table 3.

Table 2. Physical parameters of blood taken in the bioheat equation

| Tissue | $\rho_b$ [kg/m$^3$] | $C_b$ [J/(kg·K)] | $T_b$ [K] | $\omega_b$ [1/s] |
|---|---|---|---|---|
| Blood | 1020 | 3640 | 310.15 | 0.004 |

Table 3. Electrical parameters used in the simulation [10, 11]

| | $\varepsilon_r$ | $\mu_r$ | $\sigma$ [S/m] |
|---|---|---|---|
| liver tissue | 43.3 | 1 | 1.69 |
| dielectric | 2.03 | 1 | 0 |
| catheter | 2.60 | 1 | 0 |
| air slot | 1 | 1 | 1 |

Equations (16) and (21) with appropriate boundary conditions were solved using the finite element method. The simulation results are summarized in Figures 3 – 4. Fig. 3 represents the distribution of isotherms in the anal model.

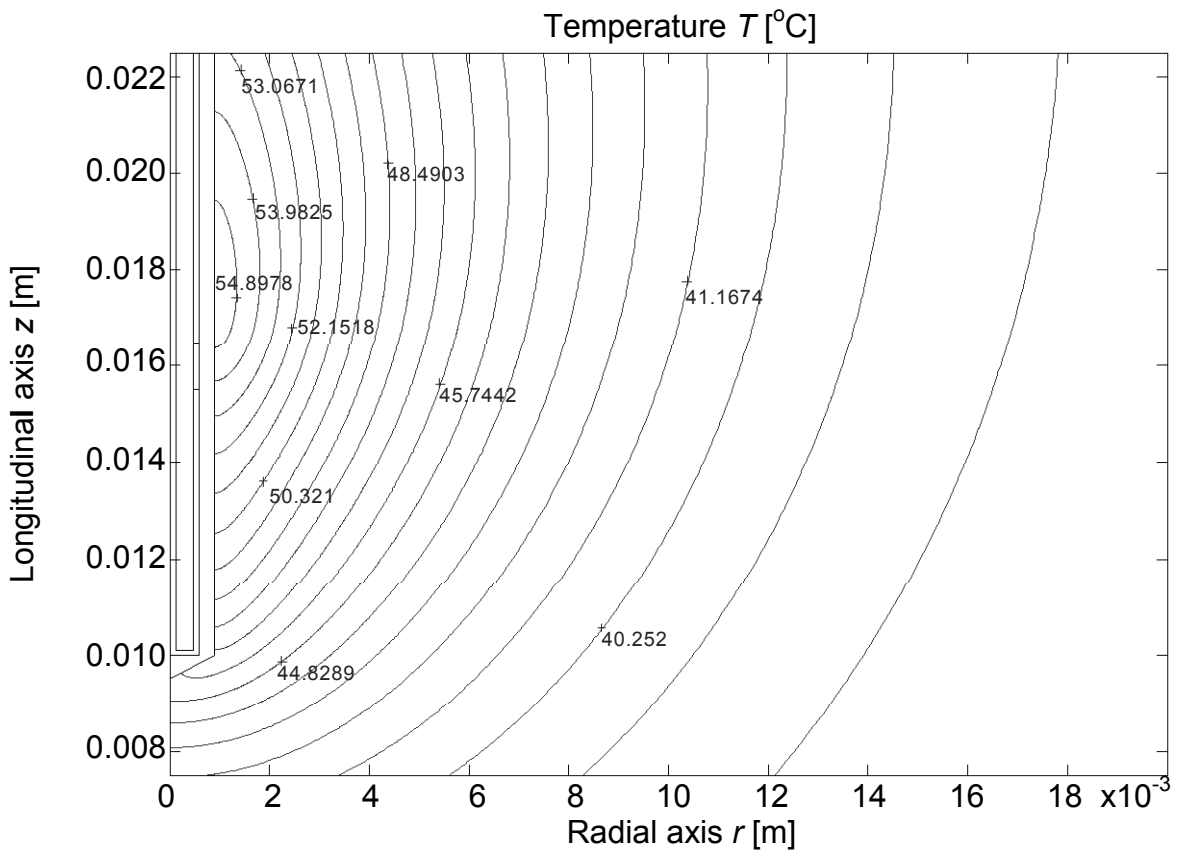

Fig.3. Isothermal lines inside the computational domain

Temperature distribution along a path crossing the tissue area at the height of the air slot is presented in Fig. 4.

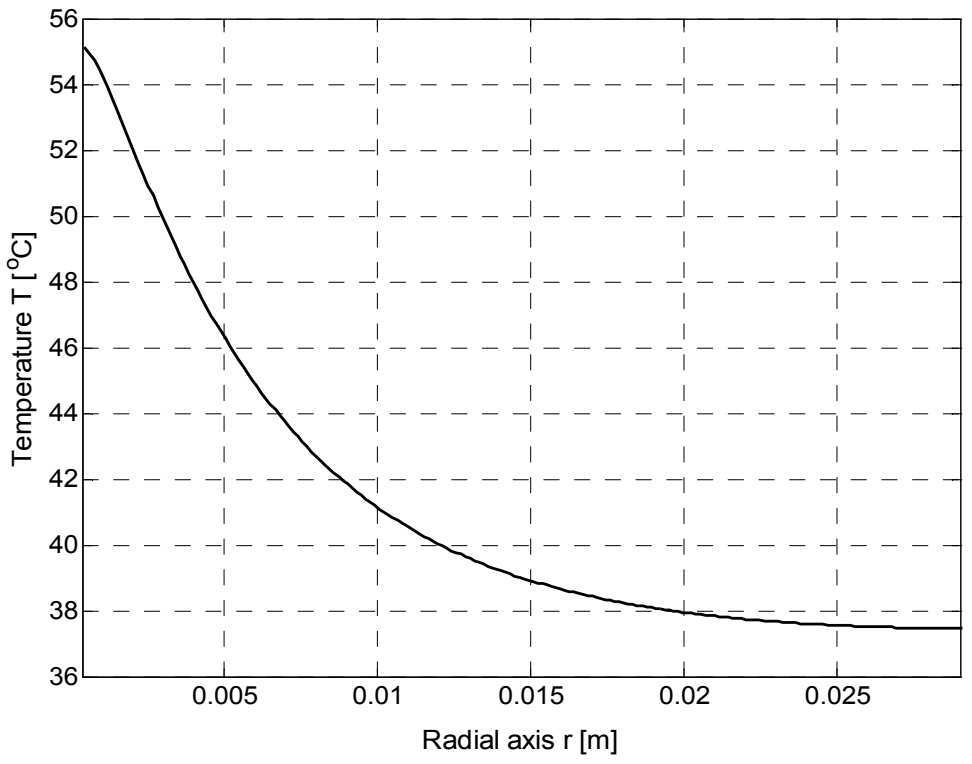

Fig.4. Temperature distribution along the path at the height of the air slot ($z$ = 0.016 m)

## Summary

Interstitial microwave thermal therapy is an invasive type of hyperthermia treatment for cancer, in which heat produced by microwaves is used to kill pathological cells associated with tumors located deep with the human body. Numerical methods are often used for dosimetric calculations for a number of important bioelectromagnetic issues. The thermal analysis of the presented problem using the finite element method allows the estimation of the temperature distribution in the specified area. Adopted a two-dimensional model represents a simplification of reality but it fully reflects the idea of treating pathological tissues using interstitial microwave hyperthermia and can be used to evaluate the actual temperature distribution in the three-dimensional case. The demonstrated plots show that the temperature decreases rapidly with the distance from the microwave applicator. Therapeutic values of temperature (above 40°C) are just 1 cm from it. This range can be easily extended by increasing the antenna input power. It is worth noting that interstitial microwave hyperthermia has reached positive clinical results in combination with radiotherapy and chemotherapy. This method is now gaining new fields of applications, e.g., in the treatment of liver, breast, kidney, bone and lung tumors.

## REFERENCES

[1] Habash R.W.Y., Bansal R., Krewski D., Alhafid H.T., Thermal Therapy, Part 2: Hyperthermia Techniques, Critical Reviews in Biomedical Engineering, 34 (2006), No. 6, 491-542.
[2] Hurter W., Reinbold F., Lorenz W.J., A Dipole Antenna for Interstitial Microwave Hyperthermia, *IEEE Trans Microwave Theory Tech.*, 39 (1991), 1048-1054.
[3] Baronzio G.F., Hager E.D., *Hyperthermia in Cancer Treatment: A Primer*, Landes Bioscience and Springer Science + Business Media, New York (2006).
[4] Lin J.C., Wang Y.J., Interstitial Microwave Antennas for Thermal Therapy, *Int. J. Hyperthermia*, 3 (1987), No. 1, 37-47.
[5] Gas P., Temperature inside Tumor as Time Function in RF Hyperthermia, *Electrical Review*, 86 (2010), No. 12, 42-45.
[6] Kurgan E., Gas P., Estimation of Temperature Distribution inside Tissues in External RF Hyperthermia, *Electrical Review*, 86 (2010), No. 01, 100-102.
[7] Miaskowski A., Sawicki B., Krawczyk A., Yamada S., The Application of Magnetic Fluid Hyperthermia to Breast Cancer Treatment, *Electrical Review*, 86 (2010), No. 12, 99-101.
[8] Gas P., Essential Facts on the History of Hyperthermia and their Connections with Electromedicine, *Electrical Review*, 87 (2011), No. 12b, 37-40.
[9] Pennes H.H., Analysis of Tissue and Arterial Blood Temperatures in the Resting Human Forearm, *J. Appl. Physiol.*, 1 (1948), No. 2, 93-122.
[10] Saito K., Taniguchi T., Yoshimura H., Ito K., Estimation of SAR Distribution of a Tip-Split Array Applicator for Microwave Coagulation Therapy Using the Finite Element Method, *IEICE Trans. Electronics*, Vol. E84-C (2001), No. 7, 948-954.
[11] Gabriel C., Gabriel S., Corthout E., The Dielectric Properties of Biological Tissues: I. Literature survey, *Phys. Med. Biol.*, 41 (1996), 2231-2249.

***Authors**: mgr inż. Piotr Gas, AGH University of Science and Technology, Department of Electrical and Power Engineering, al. Mickiewicza 30, 30-059 Krakow, E-mail: piotr.gas@agh.edu.pl*